# Dependence of Sidechain Rotamer Preference on Backbone Conformation: Relative Free Energy Calculations for Valine and Leucine

John G. Pearson and Thomas B. Woolf[1]

Correspondence[1]

Johns Hopkins University
Department of Physiology
School of Medicine
725 N. Wolfe St.
Baltimore, MD 21205

FAX: (410) 614-4436
Office: (410) 614-2643
woolf@groucho.med.jhmi.edu

Running title: Val and Leu Rotamer Libraries and their Relative Free Energy

15 manuscript pages; 6 figures (Microsoft Word on Mac)

## Abstract:

Three dimensional relative free energy calculations are used to directly calculate the dependence of the preferred sidechain rotamers for valine and leucine on the conformation of the backbone. Specifically, umbrella restrained molecular dynamics calculations are used to sample all of Ramachandran space for chi values surrounding the common rotameric states of leucine and valine. Relative free energy slices were calculated from the biased trajectories using the weighted histogram analysis method (WHAM). The slices were connected together by another set of slices perpendicular to Ramachandran space to determine the favored rotamer for a given backbone conformation. The calculated preferences are quite similar to those seen in the backbone-dependent rotamer library of Dunbrack and Karplus, despite the fact that the current calculations neglect the effects of neighboring residues. It appears likely that these calculations could be extended to calculate the optimal sidechain conformation for a peptide with known backbone conformation in the context of structure refinement and prediction.

# Introduction

The tertiary structures of proteins are frequently studied in order to better understand their functions (e.g. Creighton, 1993). Structures can be determined using diffraction techniques, NMR spectroscopy, and, to some extent, computational models. As these methods have improved, it has become evident that the structure of the protein backbone is frequently more easily resolved than the sidechains. Furthermore, analysis of existing structures indicates that only a small fraction of possible sidechain conformations are well populated. These structures correspond roughly to the trans and gauche positions of the sidechain torsions. It has thus become increasingly common for structural biologists to use rotamer libraries, derived from the PDB, in the process of structure refinement (Dobson et al., 1998; Sanchez and Sali, 1997). As a result, sidechains in newer structures are mostly in one of the common rotameric states (Ponder and Richards, 1987; Dunbrack and Karplus, 1993). However, in the absence of very precise experimental data, it is not always clear that the sidechains are in the correct rotameric state. Molecular dynamics calculations are very sensitive to starting structures with unlikely sidechain placements; it may take on the order of microseconds for the system to find the correct rotamer, while simulations are currently limited to the nanosecond timescale. Development of a reliable test of sidechain conformation in a specific structure could improve the quality for both experimental structure refinement and simulations which use the protein structures as starting points.

Although it is well recognized that conformations of individual residues, or abstracted peptide fragments, do not generally correspond to free energy minima, the conformation of the protein as a whole does (Creighton, 1993). Ideally, the calculation of a protein free energy function over all possible degrees of freedom could be performed, but this is currently not feasible. It is possible, however, to calculate free energy functions for peptides. While it is generally not possible to partition the free energy of a complex molecule into several functions of smaller systems, in the specific case of functioning globular proteins this may be feasible due to the special nature of the system (Dill, 1997). In particular, we may assume that globular proteins are not random collections of residues on a string but a small subset of sequences where the self-interaction is sufficiently favorable to allow reproducible folding to a single 'structure'. Furthermore, the structure of the protein backbone is assumed to be known, reducing the problem to *intra*-residue interactions between side chains and backbone and *inter*-residue interactions between side chains and the rest of the protein. Since residues are generally found to have standard side chain conformations – without which maintenance of a single stable structure would be improbable – the problem reduces to finding the set of conformations which avoid unfavorable *inter*-residue interactions while minimizing the *intra*-residue free energy. This approach is similar to that of Dunbrack and coworkers (Dunbrack and Karplus, 1993; Bower et al., 1997), except that in place of using prevalence of side chain rotamers in the PDB we propose calculating the free energy difference between the rotamers in determining sidechain conformation.

Multidimentional umbrella sampling may be used with the weighted histogram analysis method (WHAM) to determine the free energy as a function of a few restrained parameters (Kumar et al., 1992; Kumar et al., 1995). For the case of determining sidechain conformations in native proteins only the sidechain modes need be considered as degrees of freedom, for the backbone is assumed to be determined. Within the side chain we assume that stress upon the system, such as from an unfavorable contact, is relieved by rotation about the torsion angles as opposed to radical departure from standard bond lengths or three atom angles. Any small deformations of bond lengths or bond angles likely correlate with larger changes in torsion angle, reducing the important number of degrees of freedom and greatly simplifying calculations.

In this paper, we have calculated free energy volumes for valine and leucine as functions of backbone and sidechain dihedral angles. These volumes are compared to the empirical rotamer database of Dunbrack and co-workers (Bower et al., 1997) to see if they correctly predict the most favorable conformer for each region of Ramachandran space.

## Methods

*Model Fragments*
In principle, we could directly construct free energy hypervolumes in whatever dimensionality was appropriate to the problem. However, since WHAM calculations for more than 2 dimensions become increasingly expensive computationally, the first goal of these calculations was to cast the problem such that there were no more than 3 restrained degrees of freedom for umbrella sampling. Ideally, the selected residues should be sufficiently abundant to allow robust comparison with experimental data. Moreover, nonpolar residues are preferable, since they do not make specific inter-residue interactions such as hydrogen bonds which are neglected in our calculations. For these reasons, valiune and leucine were chosen as test residues, representing hydrophobic β- and γ- branched amino acids. Both are usually found in the protein interior, and are not involved in highly directional interactions such as hydrogen bonds.

The fragments were constructed with CHARMM, version C25, using the standard protein forcefield. Complete peptide groups were included immediately adjoining the central residue, with methyl groups blocking the ends and peptide planes on both sides of the central functional group. Figure 1 shows the chemical structures of the di-peptide fragments. Note that rotations about the $\phi$, $\psi$, and $\chi_1$ torsion angles were restricted by a 50 kcal/(mol-rad$^2$) restoring force, enabling umbrella sampling of the desired spaces. These were the only user defined restraints, with all others being part of the standard CHARMM force field (Mackerell et al., 1998).

Umbrella sampling the conformational space defined by the torsion angles $\phi$, $\psi$, and $\chi_1$ allowed generation of free energy surfaces for all three conformers of valine and for the two most common conformers of leucine, for in the absence of external contacts setting $\chi_1$ determines $\chi_2$ for these two conformers of leucine. This eliminated the need to conduct four dimensional studies of leucine conformers ($\phi$, $\psi$, $\chi_1$, $\chi_2$) or to define a set of reaction coordinates in ($\chi_1$, $\chi_2$) space at the cost of ignoring 7 sparsely populated leucine conformations.

*Molecular Dynamics*

For each residue, a series of umbrella sampled trajectories were performed, varying $\phi$ and $\psi$ on a 7 degree grid while restraining $\chi$. This yielded a total of 2601 trajectories per slice. While the exact locations of the free energy minima in $\chi$ depend on $\phi$ and $\psi$, it is reasonable to assume that they must lie near the natural rotameric states. Hence, 5 values of $\chi$ were explored for each standard conformation: $\chi_0$, $\chi_0 \pm 7$, and $\chi_0 \pm 35$, where $\chi_0$ is the standard conformation. As a result, a total of 26,010 trajectories (2 standard states * 5 $\chi$ values/standard state * 2601 $\phi/\psi$ values) were calculated for leucine, and 39,015 trajectories (3 standard states * 5 $\chi$ values/standard state * 2601 $\phi/\psi$ values) were calculated for valine. In addition, two isolated sheets of trajectories were calculated over $\psi/\chi$ space (for valine) and $\chi_1/\chi_2$ space (for leucine) to give a common reference frame for comparison of the free energies of different rotameric states. This volume slice approach (illustrated schematically for valine in Figure 2) was adopted due to the limitations of computational power available when the calculations were initiated. Sampling over the full three dimensional volume is now a reasonable but still formidable computational undertaking.

Each trajectory was generated in the following manner: an all-atom representation of the fragment was constructed from internal coordinates. A 50 kcal/(mol-rad$^2$) harmonic restraining potential was applied to restrain the $\phi$, $\psi$, and $\chi$ dihedral angles, supplementing the standard CHARMM protein forcefield. The dielectric constant was set to 80, and all nonbonded interactions were included in the calculation. The residue was relaxed by ten 50 step cycles of steepest descent minimization, followed by 5000 steps of adopted basis set Newton-Raphson minimization. This was followed by 2.5 psec of dynamics, during which structures were saved every 10 steps. All calculations were performed at 300 K with a timestep of 1 fsec. Each trajectory required approximately one minute of CPU time on a Silicon Graphics R4400 computer.

## Results

### *Free Energy Volumes*

Sets of trajectories were converted into free energy volumes by use of the Weighted Histogram Analysis Method (Kumar et al., 1992; Kumar et al., 1995). In particular we used initial code provided by Boczko and Brooks (Boczko and Brooks, 1993), modified (Serge Crouzy, CENG, France and locally) to allow determination of free energy volumes over three dimensions. Three trajectory sets near each standard conformation were converted into free energies in a single calculation, for the separation of these sets is identical to that of the grid within each set. The two outlying trajectory sets, at $\chi_0 \pm 35$, were converted into free energy surfaces in separate calculations. Overall, free energy volumes for 15 values of $\chi$ over Ramachandran space were calculated for valine (5 for each standard $\chi$ value) and for 10 values of $\chi$ over Ramachandran space were calculated for leucine (5 for each of the 2 common conformations). Additionally, two free energy surfaces in ($\psi,\chi$) space were calculated for valine and two free energy surface in ($\chi_1,\chi_2$) space were calculated for leucine. These additional free energy surfaces span spaces which intersect those at and parallel to the standard conformations, allowing selection of a common zero for free energy scales for each residue.

*Free Energy Minimum Surfaces*

We estimated the free energy minimum for each grid point in Ramachandran space from a parabolic fit of free energies from points with the same values of $\phi$ and $\psi$ from the 5 free energy volumes at and about each standard conformation. This method assumes relatively small deviations of the free energy minimum from the standard conformation, an assumption which is not necessarily correct over all of Ramachandran space. To remove artifacts generated by high energy steric contacts, free energy minimum values deemed unphysical (i.e., deviating more than 50 kcal/mol from the global minimum of each residue) or values corresponding to $\chi$ values outside the range of the fitted data (i.e., those deviating more than 35 from the standard conformation) were discarded. Values rejected from the minimal free energy volumes of valine and leucine occurred in the high energy regions of Ramachandran space, e.g. near $\phi=0$. The remaining minimal free energy values were fit with a third order Fourier series to generate analytical free energy surfaces over all of Ramachandran space, depicted in figures 3 and 4, for all three conformations of valine and the two most common conformations of leucine (Mathematica, Wolfram Research). It is important to note that while each surface is associated with a standard value of $\chi$, the actual $\chi$ value used in the surface (the one corresponding to the local free energy minimum) can vary from the standard value. In other words, each free energy surface is associated with a conformer, not a specific $\chi$ value.

*Sidechain Conformation Prediction*

Differences between the conformer free energy surfaces for a residue indicate which sidechain conformation is favored at any point in Ramachandran space. These results should correlate with the populations seen in the Protein Data Bank, assuming that intra-residue sidechain-backbone interactions are the dominant term when averaging over a large database of structures. This is equivalent to saying that inter-residue interactions, though critical to an understanding of any specific protein structure, are not well correlated with the backbone torsion angles; the inter-residue interactions are an essentially random (if large) perturbation on top of the intra-residue interactions. If this is the case, the current results should be directly comparable to those of Dunbrack and co-workers (Dunbrack and Karplus, 1993; Bower et al., 1997). We performed this comparison for valine (Figure 5) and leucine (Figure 6). The low energy regions of Ramachandran space from our calculations correspond to the highly populated regions of the empirical database, as expected. Both methods should be most accurate in these regions, the calculations because the relatively smooth surface improves WHAM convergence and the data base approach because of better statistics. Figures 5 and 6 show that the two methods generally agree on the dominant conformer, with two notable exceptions, the regions corresponding to α-helices and β-sheets. This is not surprising, since we do not explicitly represent the neighboring residues.

## Discussion:

The calculations described in this report suggest that it is now feasible to calculate relative free energy volumes for peptide fragments. The calculation of these volumes may suggest additional insights important for protein structure prediction, refinement and engineering. For example, large changes in rotamer probability from the backbone dependent free energy volumes would suggest either an error in refinement/prediction or a large contribution from inter-residue sidechain:sidechain or sidechain:mainchain interactions. This provides a key check for any protein structure: identify those points in the structure that deviate from the minimal free energy surface and then pay special attention to verify that a particular effect can explain the deviation or further refine the structure to enable the preferred rotamer to be populated.

We anticipate the extension of these calculations in several ways. Perhaps the next most obvious extension is the calculation of similar volume slices for peptide fragments that contain α-helix and β-sheet secondary structural motifs. For example, Tobias and Brooks (Tobias and Brooks, 1991; Tobias et al., 1992) performed a series of seminal free energy calculations for peptide structures to examine properties related to protein folding. We suggest that a similar series of calculations for sidechain rotamer preferences may provide insights into rotamer preferences important for structure prediction/refinement with these two common secondary structural motifs. An extension of the current calculations to a larger set of the amino acids is also anticipated. The full set of calculations will be challenging due to the presence of charged residues and the importance of explicit solvent versus implicit solvent conditions. Nonetheless, these volume free energies could provide an important rational benchmark to compare with empirical database methods and could furthermore be updated and improved with various implicit solvent models (e.g. generalized Born or multipole methods).

Certain limitations and problems with the current approach should be noted. First, since volume slices are used to approximate a calculation over the full volume, there is uncertainty introduced in the relative coupling between the volume slices. An advantage with a full volume calculation (which may soon be possible) is that no additional calibration of subvolumes to the whole is needed. An advantage of the volume slice approach is the relatively smaller amount of computer time required for each of the slices in $\chi$ space. It should also be noted that the calibration of the relative free energy between volume slices differed depending on which values of the common intersection were used for the comparison. The optimal matching between sub-volumes would occur over a larger set of intersections rather than the relatively small number of intersections in the current report. This may also be related to the assumptions of the WHAM method, in that the free energy surfaces are fully internally consistent for each WHAM sub-volume, but a full relative free energy difference between subvolumes really needs to consider all the possible (ie full volume) interactions between the points.

An additional limitation, to be noted, is that the windows used in the umbrella sampling may not be long enough for good statistical averaging. The current procedure was used to enable completion of the volume slices in a reasonable amount of computer time. A longer window for each umbrella sampled point (recall that 2061 trajectories were needed for a particular plane) adds significantly more computer time. We expect that increases in the sampling will improve the quality of the results, but will probably not change the overall trends. It would probably be better to extend the current results into the full volume space before extending the computer time allotted to each individual window. It is also significant to note that the memory requirements for the WHAM approach start to be difficult as a full volume calculation is attempted, since the currently coded algorithm requires that all data points be included in working computer memory.

The current results can be compared with conformational search approaches to the calculation of backbone dependent rotamer libraries (Dunbrack and Karplus, 1994; Marcus et al., 1996). The results from conformational search approaches provided an initial rationalization of the empirical database results. The calculations are similar to adiabatic minimized structural searches and are thus subject to the same type of errors (Brooks et al., 1988). For example, the Marcus et al calculations (Marcus et al., 1996) used the Ecepp-3 empirical energy function (Nemethy et al., 1992) and a grid search over a 20 by 20 $\phi/\psi$ range for each rotameric state of four amino acids (Met, Phe, Ile, Ser). The relative rotamer population was estimated by Boltzmann weighting the energies of each conformation, after constrained minimization. The current work extends beyond adiabatic energy surfaces to include the effects of movement at a temperature. Thus, in principle, the calculations include entropic and enthalpic degrees of freedom.

Extension of the proposed method to calculations of preferred sidechain conformations in a specific protein may be simpler than the current peptide situation. This is because the backbone conformation is set by the initial model or structure determination and the WHAM calculation can be applied in the sidechain space alone. This implies that issues of hydrogen bonding and of interactions from other regions of the protein may be sampled in a direct manner within the context of a particular protein. This would thus provide a relatively rapid test of the initial sidechain conformations determined within the model or structure determination.

An additional application of these types of calculations is to the further explanation of chemical shifts measured for protein structure determination. Currently these shifts are used heavily for assignment purposes and occasionally for checking the secondary structure type inferred through e.g. NOE and J-coupling measurements within a simulated annealing refinement approach (e.g. Evans, 1995). Recent ab initio quantum chemical calculations have suggested that an improved approach to NMR structure refinement may directly use the chemical shifts (DeDios et al., 1993; Pearson et al., 1995). An algorithm that may be possible is the Boltzmann weighting (using the relative free energy volumes) of the sidechain rotamers by backbone conformation in order to predict the probable chemical shifts for each suggested protein conformation. This may help to further extend the ab initio calculated approach to larger protein units and the application of this approach to structure refinement.

## Conclusion:

Relative free energy volumes for backbone ϕ/ψ space and sidechain χ space can now be calculated. The determination of these volumes may provide additional insights into the chemical physics of peptide structures and useful input guidance for checking the quality and improving the determination of tertiary protein structures. In particular, the surfaces cleanly separate the backbone dependent intra-residue properties of the sidechain conformations from inter-residue sidechain:sidechain and sidechain:mainchain packing interactions. Thus deviations from the calculated free energy volumes may indicate strong inter-residue interactions or suggest a problem with the structure prediction/refinement.

## Acknowledgements:

Alan Grossfield contributed substantial comments that helped with the presentation of these results. Support from the NIH (OD12600) is gratefully acknowledged.

**Figure Captions:**

Figure 1: Relative free energy volume calculations were performed using the two dipeptides depicted in parts A and B. A: The valine dipeptide is shown with torsion angles $\phi$, $\psi$, and $\chi_1$ indicated. All three torsion angles were subjected tot he umbrella sampling restraint. B: The leucine dipeptide is shown with torsion angles $\phi$, $\psi$, $\chi_1$ and $\chi_2$ indicated. The first three torsion angles were subject to the umbrella sampling restraint while $\chi_2$ was allowed to fluctuate freely.

Figure 2: Volume slices were used to perform the calculations. For both valine and leucine dipeptides, a set of trajectories (2061 in each planar slice) was used to span important regions of the full 3-dimensional volume. For example, the valine calculations illustrated in the figure consisted of seventeen trajectory sets where fifteen sets consisted of $\phi/\psi$ planar restraints over the Ramachandran space and two additional, orthogonal, calculations were performed to enable comparison across the slices. A similar calculation for leucine used twelve sets of trajectories.

Figure 3: Valine relative minimal free energy surfaces. Part A shows the contours in Ramachandran space for sidechains near $\chi$=180 degrees. Contour spacing is every 2 kcal/mol and extra contours are included 0.1 kcal/mol above and below 10 kcal/mol. Part B is a similar calculation for conformers near $\chi$=300 degrees. Part C is for $\chi$ near 60 degrees.

Figure 4: Leucine relative minimal free energy surfaces. Part A is for $\chi_1$ near 300 degrees and $\chi_2$ near 180 degrees. Contour spacing is every 2 kcal/mol with extra contours 0.1 kcal/mol above and below 10 kcal/mol. Part B is for low energy conformers near $\chi_1$ of 180 degrees and $\chi_2$ of 60 degrees.

Figure 5: Comparison of predictions for valine from relative minimal free energy surfaces versus the empirical data base of Dunbrack and co-workers (Dunbrack and Karplus, 1993; Bower et al., 1997). Dots indicate points on a ten degree grid which the WHAM relative free energy volumes suggest favor the conformation. Areas enclosed by lines indicate regions suggested by the empirical database as favoring the conformation. The shaded regions indicate at least 10 residues per 10-degree bin are included in the empirical database sample. Part A is for the $\chi_1$=180 degrees, Part B for $\chi_1$=300 degrees and Part C for $\chi_1$=60 degrees.

Figure 6: Comparison of predictions for leucine of two conformers relative to backbone $\phi/\psi$ values. Part A is the difference in relative free energies between the two conformers shown in Figure 4. The regions darker than the tripled zero line are predicted to favor the (-60, 180) conformer relative to the (180,60) conformer. Part B shows a similar plot for data from the empirical database of Dunbrack and co-workers (Dunbrack and Karplus, 1993; Bower et al., 1997). Dots indicate regions within a ten-degree bin that suggest higher population of (-60,180) relative to (180,60). Shading illustrates regions of the databank with more than 10 residues per bin.

# References

Boczko EM and Brooks CL. 1993. Constant-Temperature Free-Energy Surfaces for Physical and Chemical Processes. *Journal of Physical Chemistry* **97**(17), 4509-4513.

Bower MJ, Cohen FE, Dunbrack RL 1997. Prediction of Protein Side-Chain Rotamers From a Backbone- Dependent Rotamer Library: a New Homology Modeling Tool. *Journal of Molecular Biology* **267**(5):1268-1282.

Brooks BR, Bruccoleri RE, Olafson BD, States DJ, Saminathan S, Karplus M. 1983. CHARMM: A program for macromolecular energy minimization and dynamics calculations *J. Comp. Chem.* **4**:187-217.

Brooks CL III, Karplus M, Pettitt BM. 1988. *Proteins: A Theoretical Perspective of Dynamics, Structure, and Thermodynamics.* New York, NY: John Wiley and Sons

Creighton TE. 1993. *Proteins: Structures and Molecular Properties.* New York, NY: W.H. Freeman and Company.

Dedios AC, Pearson JG, Oldfield E. 1993. Secondary and Tertiary Structural Effects on Protein NMR Chemical-Shifts - an Ab initio Approach. *Science* **260**(5113):1491-1496.

Dill KA. Additivity Principles in Biochemistry. 1997. *Journal of Biological Chemistry* **272**(2):701-704.

Dobson CM, Sali A, Karplus M. 1998. Protein Folding: a Perspective From Theory and Experiment. *Angewandte Chemie-International Edition in English* **37**(7):868-893.

Dunbrack RL and Karplus M. 1993. Backbone-Dependent Rotamer Library for Proteins - Application to Side-Chain Prediction*Journal of Molecular Biology* **230**(2):543-574.

Dunbrack RL and Karplus M. 1994. Conformational-Analysis of the Backbone-Dependent Rotamer Preferences of Protein Side-Chains. *Nature Structural Biology* **1**(5):334-340.

Evans JNS. 1995. *Biomolecular NMR Spectroscopy.* Oxford, UK: Oxford University Press.

Kumar S, Bouzida D, Swendsen, RH, Kollman, PA, Rosenberg, JM. 1992. The Weighted Histogram Analysis Method for Free-Energy Calculations on Biomolecules .1. The Method. *Journal of Computational Chemistry* **13**(8):1011-1021.

Kumar S, Rosenberg, JM, Bouzida D, Swendsen RH, Kollman PA. 1995. Multidimensional Free-Energy Calculations Using the Weighted Histogram Analysis Method*Journal of Computational Chemistry* **16**(11):1339-1350.

Mackerell AD, Bashford D, Bellott M, Dunbrack RL, Evanseck JD, Field MJ, Fischer S, Gao J, Guo H, Ha S, Joseph-Mccarthy D, Kuchnir L, Kuczera, K, Lau FTK, Mattos C, Michnick S, Ngo T, Nguyen DT, Prodhom B, Reiher WE, Roux B, Schlenkrich M, Smith JC, Stote R, Straub J, Watanabe, M, Wiorkiewicz-Kuczera J, Yin D, Karplus M. 1998. All-Atom Empirical Potential for Molecular Modeling and Dynamics Studies of Proteins*Journal of Physical Chemistry B* **:** **102**(18):3586-3616.

Marcus E, Keller DA, Shibata M, Ornstein RL, Rein R. 1996. Comparing Theoretical and Experimental Backbone-Dependent Sidechain Conformational Preferences for Linear, Branched, Aromatic and Polar Residues. *Chemical Physics* **204**(2-3):157-171.

Nemethy G, Gibson KD, Palmer KA, Yoon CN, Paterlini G, Zagari A, Rumsey S, Scherga HA. 1992. Energy Parameters in Polypeptides. 10. Improved geometrical parameters and nonbonded interactions for use in the ECEPP/3 Algorithm, with Application to Proline-Containing Peptides. Journal of Physical Chemistry **96**(15):6472-6484.

Pearson JG, Wang JF, Markley JL, Le HB, Oldfield, E. 1995. Protein-Structure Refinement Using C-13 Nuclear-Magnetic- Resonance Spectroscopic Chemical-Shifts and Quantum-Chemistry. *Journal of the American Chemical Society* **.** **117**(34):8823-8829.

Ponder JW and Richards FM. 1987. Tertiary templates for proteins: Use of packing criteria in the enumeration of allowed sequences for different structural classes *J. Mol. Biol.* **193**:775-791.

Sanchez R and Sali A. 1997. Comparative Protein Structure Modeling as an Optimization Problem. *Theochem-Journal of Molecular Structure* **398**:489-496.

Tobias DJ, Brooks CL III. 1991. Thermodynamics and mechanism of alpha-helix initiation in alanine and valine peptides. *Biochemistry* **30**:6059-6070.

Tobias DJ, Sneddon, SF, Brooks CL III. 1992. Stability of a Model Beta-Sheet in Water. *Journal of Molecular Biology* **227**(4):1244-1252.

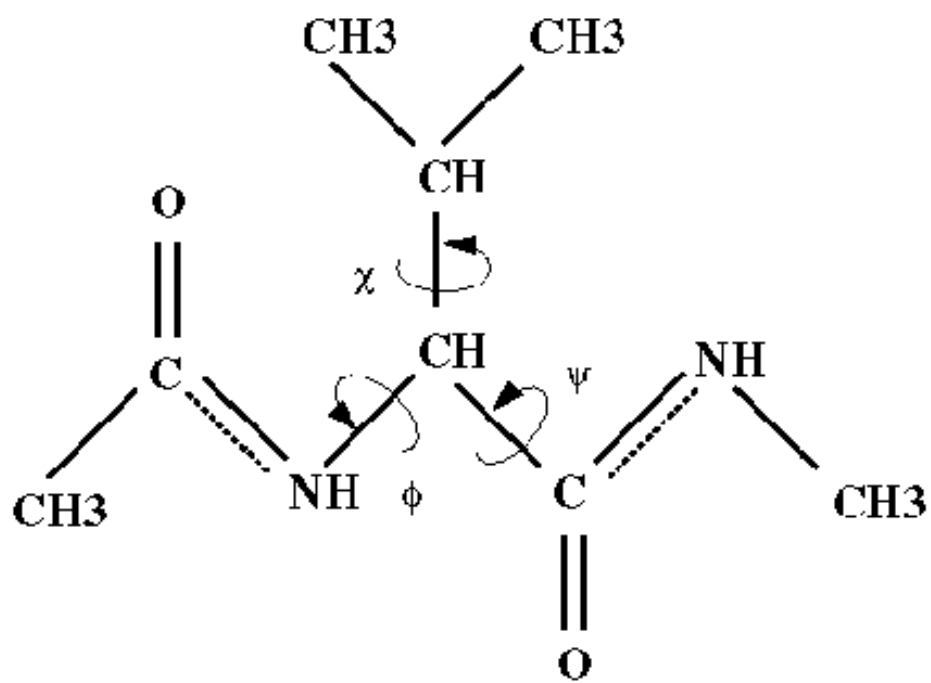

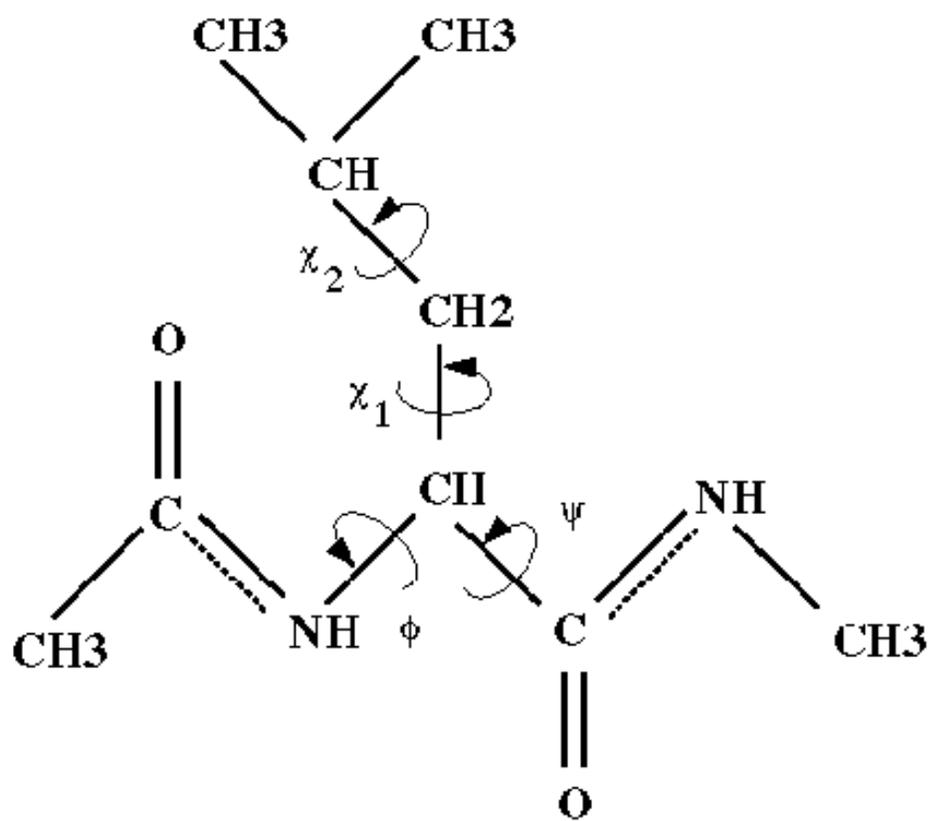

Fig 1

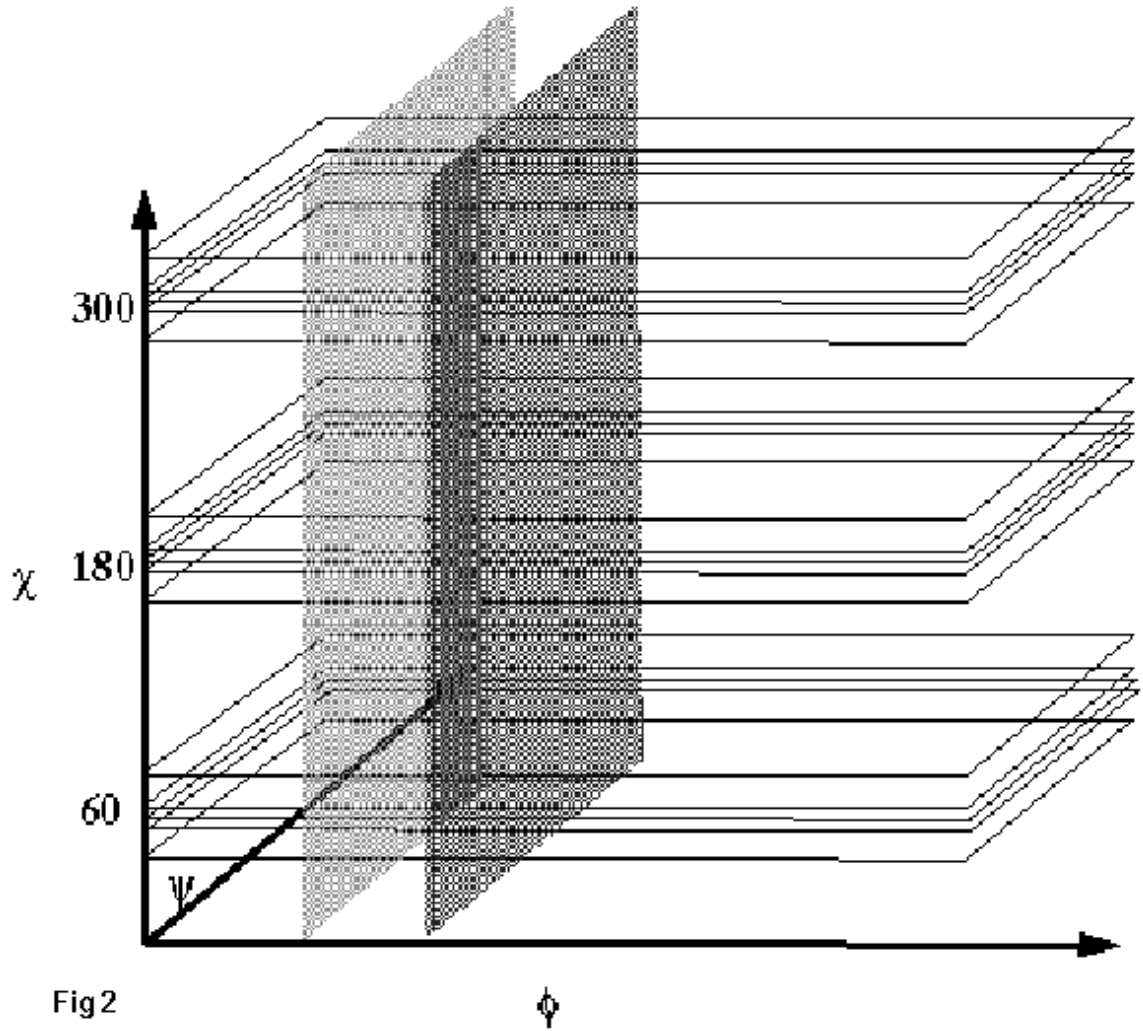
Fig 2

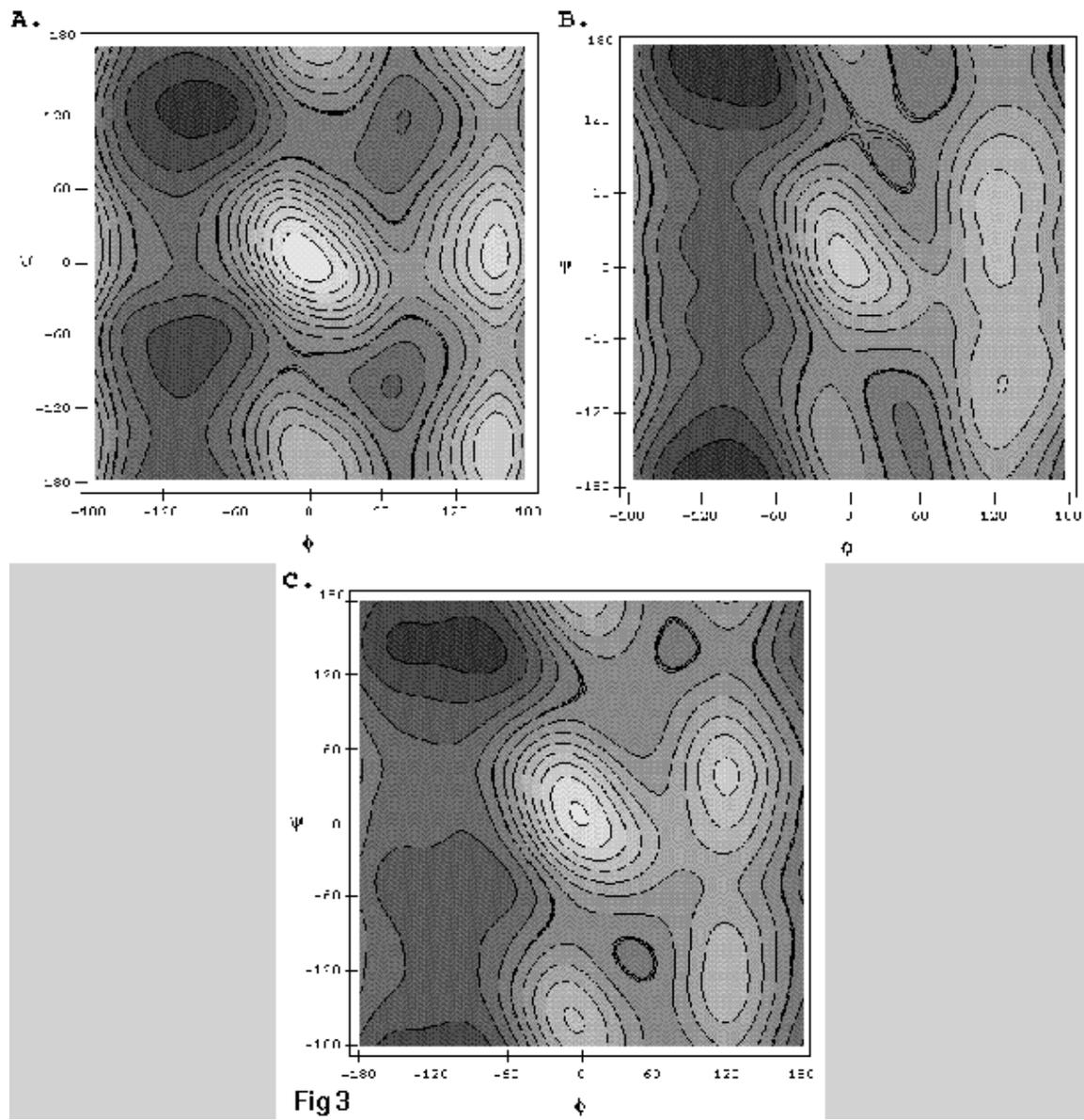
Fig 3

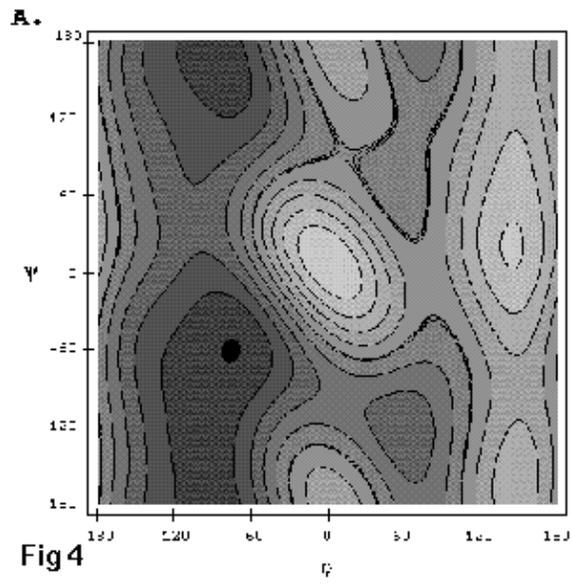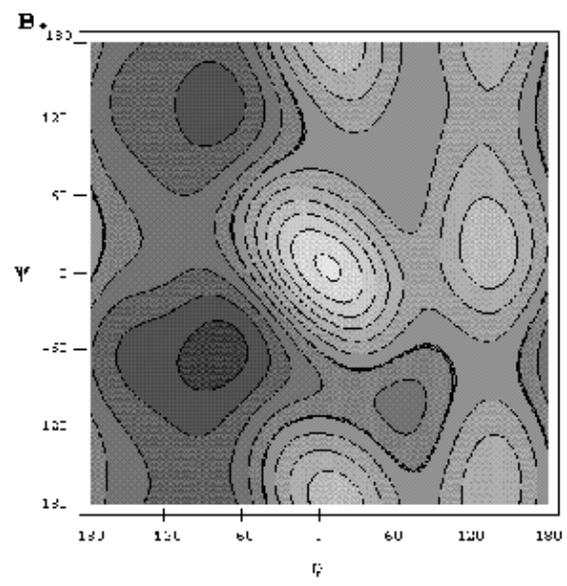

Fig 4

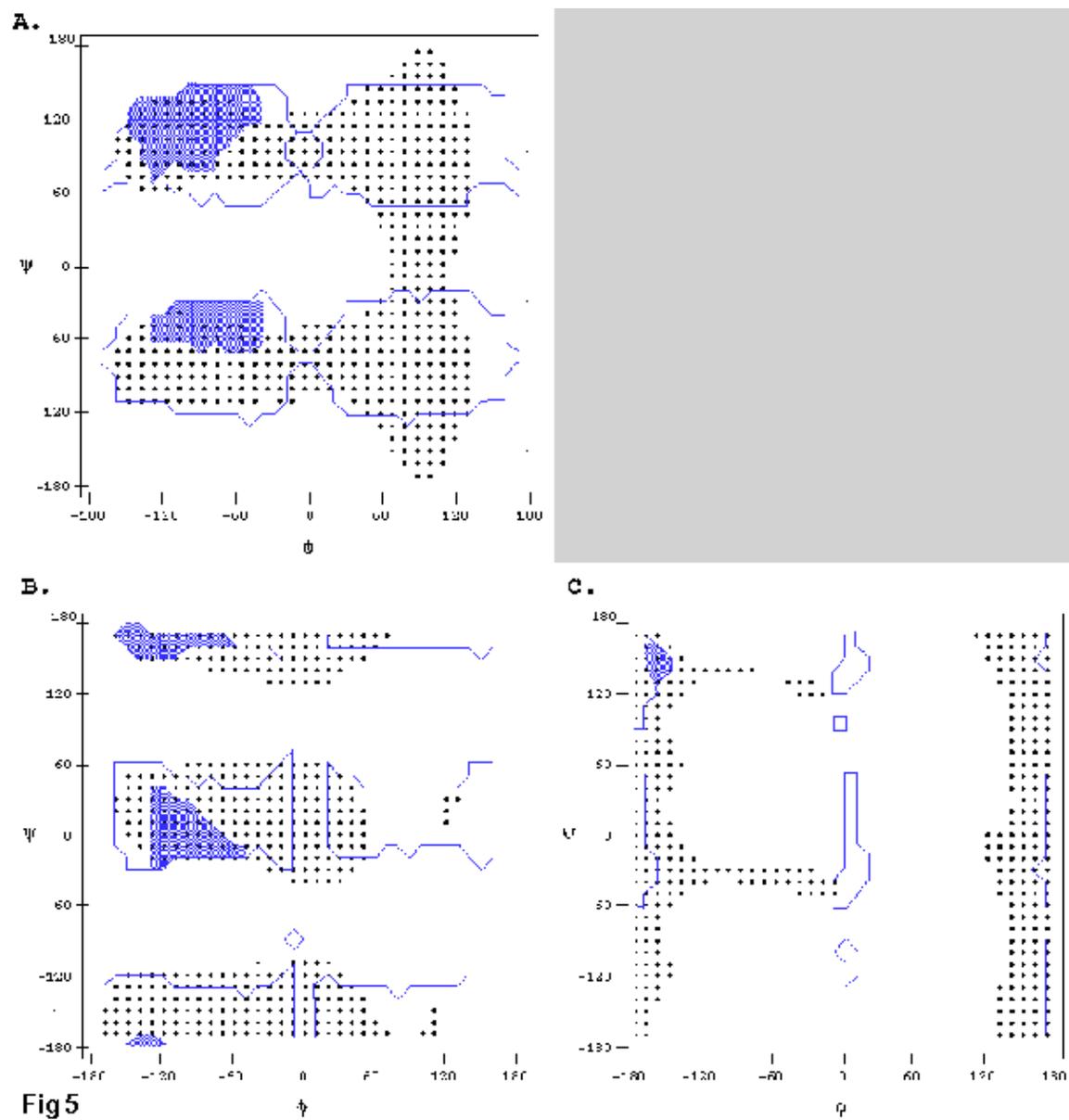

Fig 5

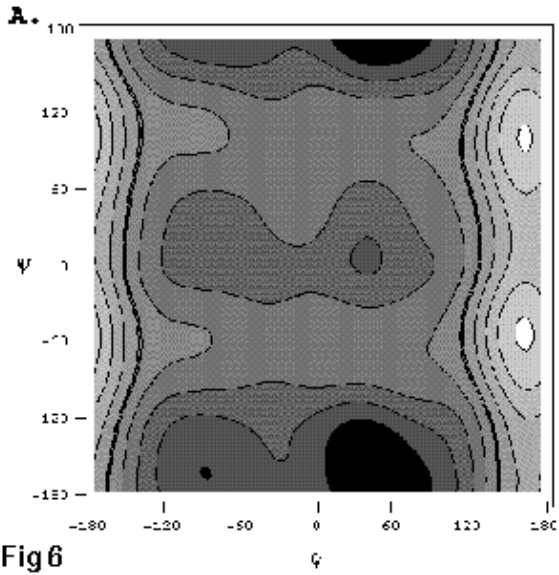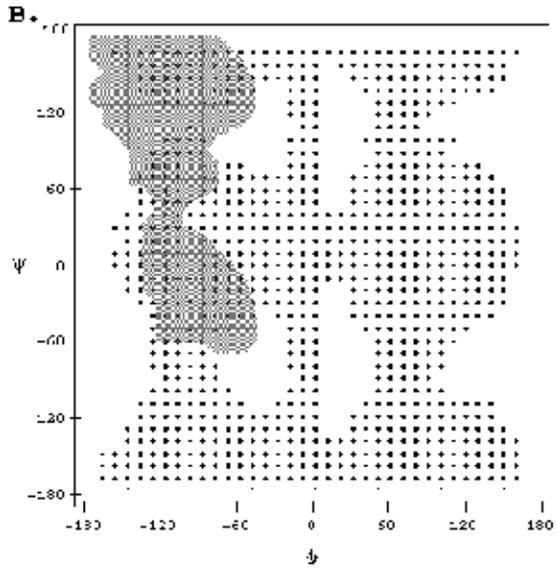

Fig 6